\documentclass[a4paper,11pt]{article}
\pdfoutput=1
\usepackage{jcappub}
\usepackage{hyperref}
\usepackage[T1]{fontenc} 

\title{A novel double-rim forebaffle design for centimeter to sub-millimeter astrophysical observations}

\author[a,b,1]{Jacques Delabrouille,\note{Corresponding authors.}}
\author[a,b,1]{Oliver Jeong,}
\author[c,a,b]{Michel Piat}
\author[b]{and Alexander Steier}

\affiliation[a]{CNRS-UCB International Research Laboratory, Centre Pierre Bin\'etruy, \\
IRL 2007, CPB-IN2P3, Berkeley, CA 94720, USA}
\affiliation[b]{Lawrence Berkeley National Laboratory, \\
1 Cyclotron Road, Berkeley, CA 94720, USA}
\affiliation[c]{Universit\'e Paris Cit\'e, CNRS, Astroparticule et Cosmologie, \\
F-75013 Paris, France}
\emailAdd{delabrouille@apc.in2p3.fr}
\emailAdd{objeong@lbl.gov}
\emailAdd{piat@apc.in2p3.fr}
\emailAdd{a.steier@campus.lmu.de}

\abstract{Stray radiation of various origin is a major source of degradation of centimeter to sub-millimeter astronomical observations. This is particularly problematic for the detection of signals such as faint cosmic microwave background polarization $B$ modes, or for mapping large-scale extragalactic or Galactic diffuse emission. In this paper, we propose a double-rim forebaffle design to reduce the impact of such stray radiation contamination. Using qualitative arguments and numerical simulations, we show that such a design has the potential to substantially improve the quality of future observations.}

\begin{document}
\maketitle
\flushbottom

\section{Introduction}
Astronomical observations at centimeter to sub-millimeter wavelengths are key to probing the distant and cold universe. Target emissions include the Cosmic Microwave Background (CMB); the Sunyaev-Zel'dovich (SZ) effect towards galaxy clusters; the distribution of line emissions from distant galaxies used to probe the three-dimensional distribution of matter with millimeter-wave line intensity mapping (LIM); and emissions by the interstellar medium (ISM) of our own Galaxy, the Milky Way.

When ground-based telescopes scan the sky to map centimeter to sub-millimeter emissions, detectors also measure spurious signals from unwanted sources. Among those, stray radiation from the ground, nearby buildings and mountains, and celestial objects such as the Sun, the Moon, and the Milky Way are of particular concern. Avoiding excessive contamination constrains the range of azimuth and elevation allowed during the observations to ensure that sources of stray radiation are kept well away from the line of sight of the telescopes. 
These scanning constraints reduce the global observing efficiency, and residual contamination degrades the quality of the observations. The data-streams are often high-pass filtered to minimize contamination, resulting after their projection onto maps in anisotropic and non-stationary filtering of the signals of interest.

To minimize this observational non-ideality, it is essential to design telescopes that are minimally susceptible to contamination by stray signals originating from unwanted sources of radiation on Earth and in the sky. This is traditionally achieved with a set of shields and baffles that screen focal-plane detectors from direct illumination by sources of unwanted incoming radiative power. 
In this paper, we introduce a novel co-moving double-rim forebaffle concept that has the potential to substantially reduce the contamination of the observations by stray radiation as compared to some of the current forebaffle solutions. We motivate the design by the need to minimize the impact of different types of stray radiation sources, and specifically demonstrate substantial suppression of diffracted sidelobe power from nearby mountains in the case of observations near Cerro Toco on the Atacama plateau.

The remainder of this paper is organized as follows. In Section~\ref{sec:problem}, we review the main sources of stray radiation for a small refracting telescope and traditional methods to minimize their amplitude.
Section~\ref{sec:forebaffle} introduces the double-rim forebaffle and discusses design options as a function of observation conditions. 
In Section~\ref{sec:sidelobes}, we use a far-sidelobe modeling tool introduced by \cite{jeong} 
to compare and discuss expected sidelobe levels for a standard single-rim forebaffle and the new double-rim forebaffle design. 
We conclude in Section~\ref{sec:conclusion}.

\section{Straylight in small aperture CMB telescopes}
\label{sec:problem}

As a concrete example, we consider small refractive telescopes designed to detect primordial, parity-odd $B$~modes of CMB polarization at angular scales of about a degree or larger, looking for direct confirmation that the early universe experienced a phase of very rapid expansion, called cosmic inflation \cite{2016ARA&A..54..227K}. However, most of our discussion is applicable to ground-based telescopes designed for a range of potential observations at centimeter to sub-millimeter wavelengths.

\begin{figure}[htbp]
\centering\includegraphics[width=0.75\textwidth]{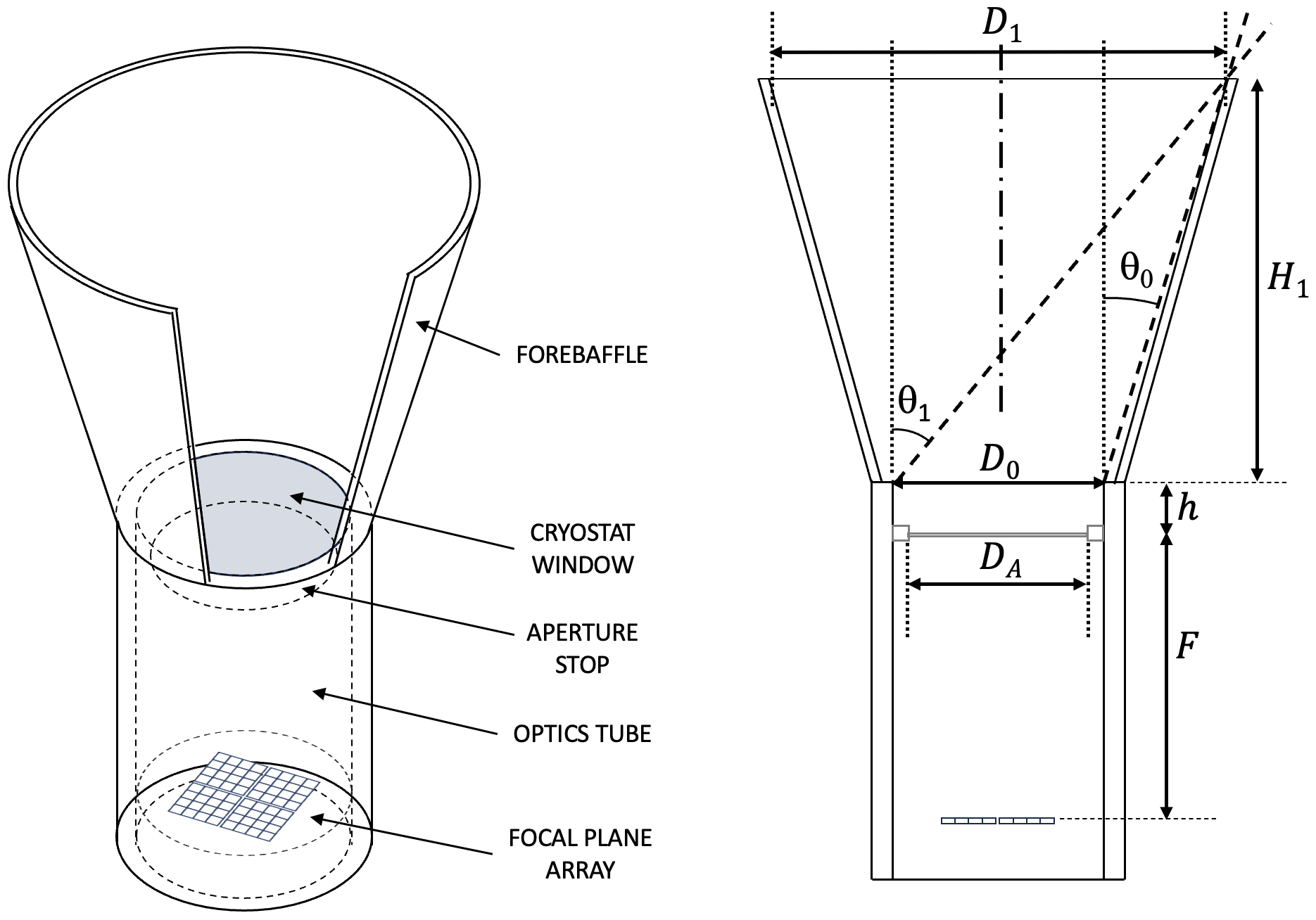}
\caption{Sketch of the optics tube and forebaffle of a ground-based refractive small-aperture telescope designed for the detection of CMB polarization $B$~modes. Left: three-dimensional view, with a section of the forebaffle cut out; Right: cut through a plane of symmetry. The diameters of the entrance aperture and of the rim of the forebaffle are denoted  $D_0$ and $D_1$ respectively, $F$ is the focal length and $H_1$ is the height of the forebaffle. These dimensions define the angles $\theta_0 = \arctan((D_1\!-\!D_0)/2H_1)$ and $\theta_1 = \arctan((D_1\!+\!D_0)/2H_1)$. 
Optical elements such as lenses and filters are not represented in this figure.}
\label{fig:SAT-geometry}
\end{figure}

Typical telescopes designed for the detection of large-scale CMB polarization $B$ modes, such as the BICEP3 telescope \cite{2018SPIE10708E..2NK}, Simons Observatory small aperture telescopes \cite{2024ApJS..274...33G}, and the AliCPT instrument \cite{2020SPIE11453E..2AS} comprise a main optics tube with a cold entrance aperture, lenses and filters, and a focal plane populated by cryogenic transition-edge sensors or kinetic inductance detectors. 
On top of the optics tube, a co-moving forebaffle protects the entrance aperture from direct radiation originating from any bright terrestrial or celestial source of radiation away from the sky region of interest. A representative telescope comprised of an optics tube and a forebaffle is sketched in Figure~\ref{fig:SAT-geometry}.

For small-aperture telescopes, a ground shield additionally protects the detectors from direct illumination by the ground and building nearby, but this shield may not always be sufficient to cover all of the terrestrial environment (see section \ref{sec:sidelobes} for a practical case study). 

\subsection{The stray radiation problem}

\begin{figure}[htbp]
\centering\includegraphics[width=\textwidth]{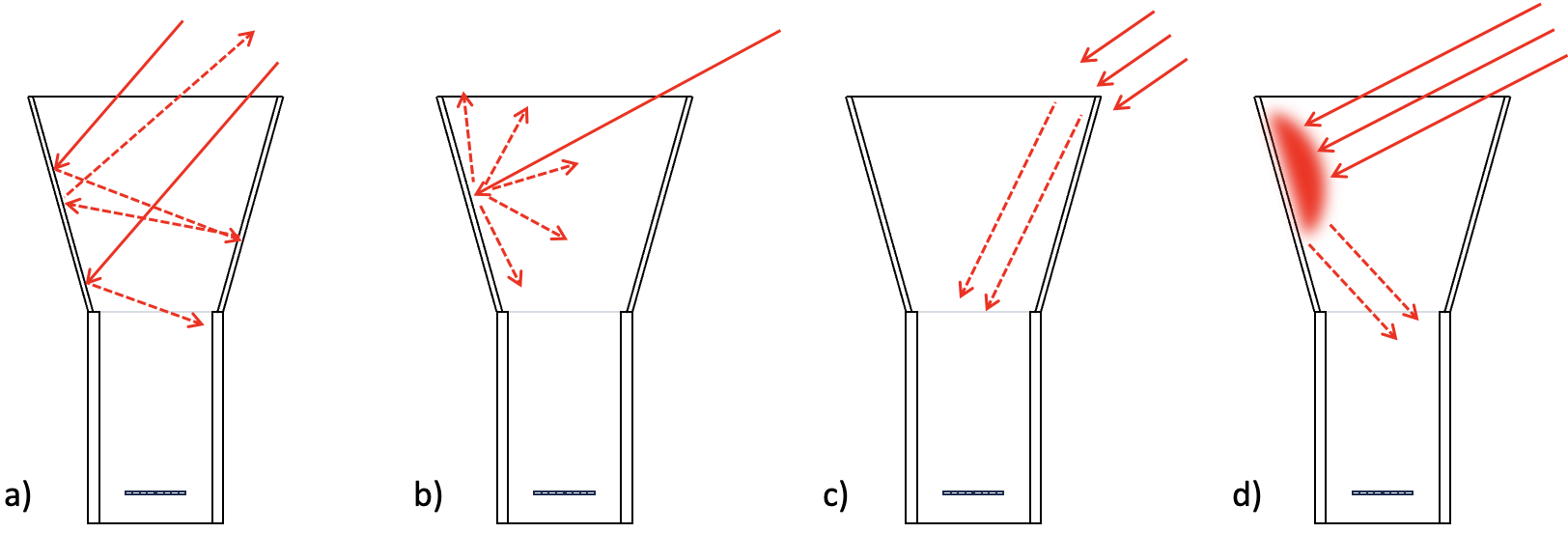}
\caption{Illustration of four mechanisms for straylight pickup by a 
telescope comprised of a refractive optics tube and a forebaffle. a) Reflections of radiation inside the forebaffle; b) Scattering by the inner surface of the forebaffle; c) Diffraction by the forebaffle rim; d) Thermal radiation from the inner forebaffle heated by the Sun. }
\label{fig:straylight}
\end{figure}

Detectors in the focal planes of CMB telescopes are sensitive to any source of incoming radiation. As the 
telescope scans the sky according to the observing strategy, data streams are formed that contain the main signal of interest. However, these data streams also register the integral of the fluctuations due to incoming emissions from other origins than the main detector beams.
Stray radiation enters the instrument and contaminates the observations through four main processes, illustrated in Figure~\ref{fig:straylight}: 
    a)~Specular reflection of off-axis light rays by reflective surfaces, away from the nominal main-beam rays, causing ``ghost beams'' or strong sidelobes in specific directions;
    b)~Scattering of microwave radiation by surfaces that are optically coupled to the detectors, such as the entrance aperture, filters, and the inner surface of the forebaffle;
    c)~Diffraction of microwave radiation by the rims of the shields, baffles, and the entrance window;
    d)~Thermal emission from the instrument itself, originating from the absorption of sunlight by instrument surfaces and re-emission at CMB wavelengths towards the optics tube.

All of these are potentially problematic when strong sources of radiation, such as the Sun, the Moon, or large terrestrial horizon features such as mountains and buildings, illuminate the inner surface of the forebaffle. Direct solar illumination inside the forebaffle potentially generates all of the above four kinds of stray radiation pick-up. 
In the following, we discuss each of those processes in the specific context of a refractive small-aperture telescope (SAT) optics tube with a single conical or cylindrical forebaffle, similar to the optics tubes of BICEP3, the Simons Observatory, and AliCPT. We focus our discussion on straylight propagation between the forebaffle rim and the optics tube entrance aperture.

\subsection{Reflections}
Light rays that directly illuminate the interior of the forebaffle can be reflected by the inner surface and bounce around to reach the entrance window of the optics tube. 
A conical forebaffle reflects out some of the light that illuminates the inner surface, but not all of it. When the inner surface of the conical forebaffle is defined by an entrance window of diameter $D_0$ and a top aperture of diameter $D_1$, with an opening angle $\theta_0 = \arctan((D_1-D_0)/2H_1)$, 
it can be shown \cite{Winston2005} that all rays with incident angle higher than $\theta_0 + \arctan(D_0/D_1)$ are always reflected back to the sky. 
Besides optimizing the geometry, straylight via reflection can also be reduced by coating the inner surface of the forebaffle with an absorbing material, such as eccosorb at millimeter and centimeter wavelengths or black paint at optical wavelengths. The interior of the forebaffle can also be corrugated with platelets.
In practice, a fraction of the incident radiation will always be reflected inside the forebaffle and make it to the entrance window, but this fraction can be minimized by design to a level ranging from a few percent to a fraction of a percent.

\subsection{Scattering}
If the inner surface of the forebaffle is not perfectly absorbing, some of the incident light can also be scattered toward the optics tube entrance window and eventually reach the detectors. 
Stray radiation via scattering from the forebaffle can be mitigated with high-quality absorbing material. However, surfaces are never perfectly absorbing and can also be contaminated by deposited scatterers such as dust particles or ice, so a small fraction of the incident radiation always reaches the focal plane. Reducing the total illumination of the inner forebaffle is a practical way to reduce this effect.

\subsection{Diffraction}
Diffraction
is an unavoidable source of stray radiation in astronomical telescopes. At centimeter to sub-millimeter wavelengths, it is one of the main sources of off-axis radiation pick-up. Ray-tracing or physical optics software 
can be used to compute the antenna patterns of the telescopes and evaluate the amplitude of sidelobe signals during the observations. However, accurate calculations require a precise definition of the geometry of the entire instrument, including surface properties. This is possible only at some level of approximation. Main beams and near-sidelobes can be accurately evaluated, but far-sidelobe levels are often uncertain. 
Calculations offer qualitative or comparative assessments, but cannot be used in practice to evaluate and subtract very-far-sidelobe signals from the data streams. 

\subsection{Thermal emission straylight}

If the Sun is allowed to directly illuminate the inner surfaces of the forebaffle during the observations, solar rays will be absorbed and heat those inner surfaces in a time-dependent way. Energy will be re-radiated in the microwave wavelength range and enter the optics tube through its entrance window. 
Note that the Sun also heats the outer surface of the forebaffle during the scan, but a good thermal insulator between the inner and outer surfaces will keep the thermal fluctuations of the outer surface from propagating to the inner surface that is directly visible from the entrance window.

\section{The double-rim forebaffle design}
\label{sec:forebaffle}

All of these sources of straylight can be reduced by shielding the inner forebaffle and its rim from direct view of strong radiation sources. We now introduce a novel double-rim forebaffle design that avoids direct illumination of the inner forebaffle and rejects straylight originating from boresight angles larger than some target rejection angle $\theta_2$. 

\begin{figure}[htbp]
\centering\includegraphics[width=0.9\textwidth]{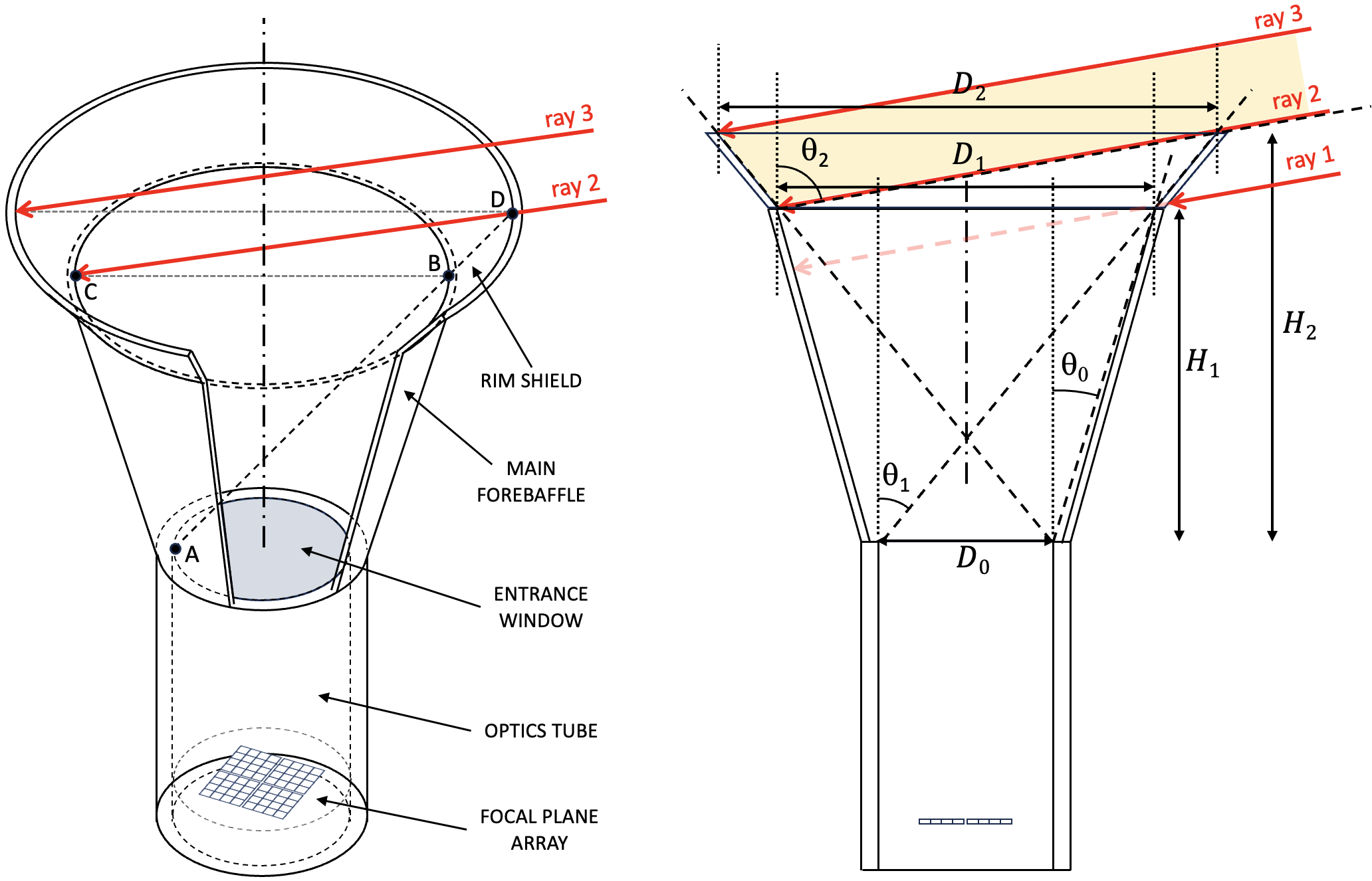}
\caption{The double-rim forebaffle concept. The geometry defines three angles of interest: $\theta_0$ is the limit below which all light rays intercept the entrance window; $\theta_1$ is the limit above which no light ray reach the entrance window; $\theta_2$ is the limit above which no light ray reach the inner main forebaffle. The slanted rim shield intercepts light rays between ray~1 and ray~2, which otherwise would illuminate the inner forebaffle. When the solar aspect angle $\theta_{\rm Sun}$ satisfies $90^\circ > \theta_{\rm Sun} > \theta_2$, the Sun shines only on the inner surface of the rim shield, and no reflected, scattered, or re-emitted light ray points directly towards the entrance window.
The yellow-shaded area illustrates the bundle of rays that illuminate the rim shield when the Sun is at an angle $\theta_{\rm Sun} = \theta_2$.}
\label{fig:flowerpot}
\end{figure}

\subsection{Conceptual guidelines}

The key idea is to reduce the inner forebaffle area that is i) illuminated by strong sources of radiation (the Sun in particular); and ii) directly visible from the optics tube entrance window.
The concept is illustrated in Figure~\ref{fig:flowerpot}. An extra baffle, the ``rim shield,'' is added on top of the main forebaffle to protect the inner surfaces of the main forebaffle from direct radiation incoming from angles above some rejection angle $\theta_{\rm 2}$. The geometry of the full system defines three angles of interest.
The first angle, $\theta_0$, is the angle from the boresight below which we want \emph{all} rays to reach the entrance window of the optics tube. This angle is defined by the required telescope FOV. Next, $\theta_1$ is the limit angle above which we want \emph{no} light ray to reach the entrance aperture directly. This could be, for instance, the minimal lunar avoidance angle. Finally, $\theta_2$ is the rejection angle above which we want no light ray to directly reach the inner surfaces of the main forebaffle, and hence diffracts or scatters \emph{twice} to reach the entrance window. This could be, for instance, the required minimum solar or ground avoidance angle.

The shadowing provided by the rim shield depends only on the diameter $D_2$ of its top aperture and on its distance $H_2$ from the entrance aperture. There is freedom to define the exact shape of the surface of the rim shield. However, 
a conical shape that is as open as possible scatters, reflects, or re-radiates incoming light rays maximally away from the entrance aperture. This leads to the geometry of the double cone ``flowerpot-like'' forebaffle sketched in Figure~\ref{fig:flowerpot}.

\subsection{Double-rim forebaffle dimensioning prescriptions and toolkit}

We now turn to practical designs for the double-rim forebaffle, starting from existing forebaffles of experiments dedicated to the detection of CMB primordial polarization $B$ modes. We consider three cases, similar to the following telescopes: i- BICEP3 located at the South Pole \cite{2018SPIE10708E..2NK}; ii- Simons Observatory SATs located in Chile \cite{2024ApJS..274...33G}; iii- The AliCPT-1 forebaffle designed by Beijing's Institute of High Energy Physics (IHEP) team for the AliCPT-1 instrument located in Tibet \cite{2020SPIE11453E..2AS}. The dimensions of each of them are listed in Table \ref{tab:3-SAT-designs}.

\begin{table}[htbp]
\centering
\begin{tabular}{|l|ccccccc|}
\hline
Telescope & $D_A$ & $D_0$ & $D_1$ & $H_1$ & FOV diameter & $\theta_0$& $\theta_1$ \\
\hline
BICEP3 & 52~cm & 73~cm & 130~cm & 130 cm & $27.4^\circ$ & $12.4^\circ$ & $38.0^\circ$ \\
SO SATs & 42~cm & 69~cm & 210~cm & 170 cm & $35.0^\circ$ & $22.5^\circ$ & $39.4^\circ$ \\
AliCPT-1 & 72~cm & 97~cm & 183.6~cm & 199~cm & $20.8^\circ$ & $12.3^\circ$
& $35.2^\circ$ \\
\hline
\end{tabular}
\caption{Dimensions of existing refractive telescope forebaffles for the detection of primary CMB $B$ modes. As we do not have access to exact design and construction data for all of these forebaffles, which also may evolve during the lifetime of those experiments, numbers should be considered as indicative only.}
\label{tab:3-SAT-designs}
\end{table}

\begin{figure}[htbp]
\centering\includegraphics[width=0.7\textwidth]{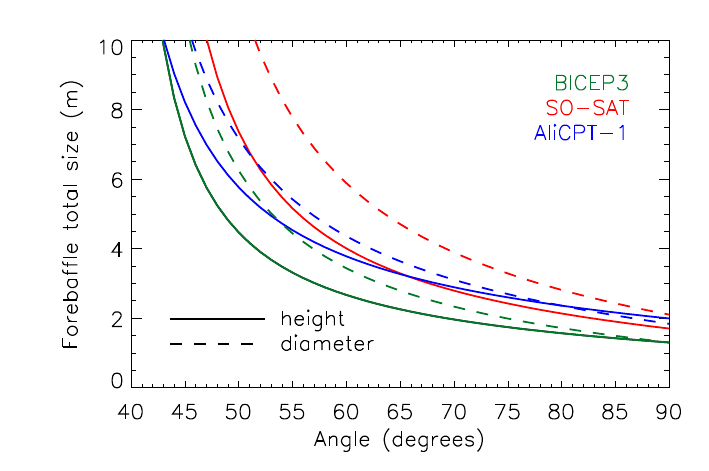}
\caption{Total double-rim forebaffle height and diameter, as a function of required rejection angle, for three different forebaffle designs as specified in table~\ref{tab:3-SAT-designs}.}
\label{fig:plot-D2H2}
\end{figure}

Given these forebaffles, we compute the minimum dimensions of the rim shield for each configuration and for a set of required rejection angles $\theta_2$ ranging from $40^\circ$ to $90^\circ$. The minimum vertical extent $H_2$ and diameter $D_2$ of the rim shield are obtained from the following set of equations:
\begin{eqnarray}
     2H_2 \tan\theta_1 & = & (D_0\!+\!D_2) \label{eq:h2d2-1}\\
     2(H_2\!-\!H_1) \tan\theta_2 & = & (D_1\!+\!D_2) ,
\end{eqnarray}
from which we get
\begin{equation}
    H_2 = (D_0\!-\!D_1\!-2 H_1 \tan \theta_2) / (\tan \theta_1 \! -\! \tan \theta_2) \label{eq:h2}
\end{equation}
and we get $D_2$ by plugging the value for $H_2$ into Equation~\ref{eq:h2d2-1}.
These dimensions are plotted in Figure~\ref{fig:plot-D2H2}.
As expected, shielding rays from smaller aspect angles requires increasingly large rim shields as we tighten the constraint on $\theta_2$. Practical dimensioning requires balancing the added value of the rim shield in terms of straylight rejection against the engineering complexity and size of the entire system. 

We provide a Desmos graphing calculator page\footnote{\url{https://www.desmos.com/calculator/u64ts2du58}} that can be used to create a simple 2-D model of the double-rim forebaffle, in which one can freely alter instrument dimensions and see the effect it has on the size of the rim shield. This implements Equations~\ref{eq:h2d2-1} to \ref{eq:h2} and also provides toggleable sliders for each dimension of the instrumental setup.

\section{Diffraction sidelobe profile calculations and comparisons}
\label{sec:sidelobes}

To illustrate the benefits of a double-rim design, we calculate the diffraction sidelobe profile of an example forebaffle for the BICEP3 telescope, showing the comparative suppression of far-sidelobe power with and without this component using parameters listed in Table~\ref{tab:3-SAT-designs} and a rejection angle of $\theta_2=60^{\circ}$. The beam profile is modeled using the method described in \cite{jeong}. This method uses \textsc{Diffractio}~\cite{diffractio}, an open-source Python package based on the principles of the geometric theory of diffraction, to model the propagation of light through free space and diffracting apertures. The telescope is simplified into a series of diffracting apertures. We model the focal plane excitation by defining a detector's distance and angle from the center of the entrance aperture for a specified f-number and wavelength of light. This approach produces a Gaussian wave that propagates at the specified angle out of the entrance aperture with the corresponding beam waist. For more detail, refer to \cite{jeong}.
\par
\begin{figure}[htbp]
\centering\includegraphics[width=0.7\textwidth]{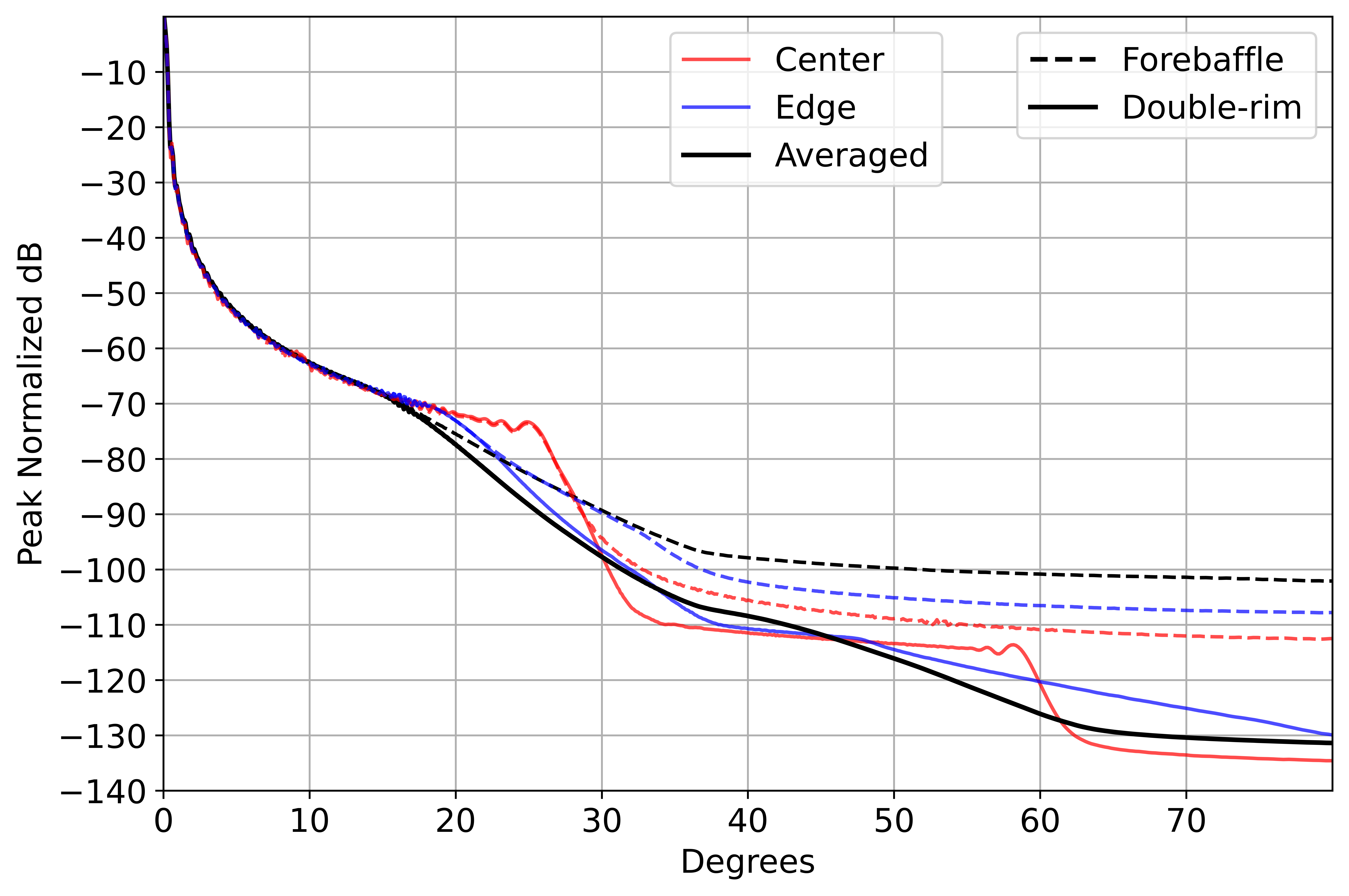}
\caption{Peak normalized beam profile of a center pixel (red), edge pixel (blue), and averaged over the focal-plane (black) for traditional forebaffle (dashed) and double-rim forebaffle (solid) designs. Note the double-rim forebaffle with a 60$^{\circ}$ rejection angle shows a strong suppression in sidelobe power at 60$^{\circ}$.}
\label{fig:beam_prof}
\end{figure}

Figure~\ref{fig:beam_prof} presents the results of beam profile calculations for BICEP3, both with just a regular forebaffle, and with a double-rim forebaffle with a rejection angle of $\theta_2=60^{\circ}$. All profiles shown are azimuthally-averaged and band-averaged. We show the profile averaged over the focal plane, as well as the profiles for the center pixel and a pixel at the furthest edge of the focal plane. 
The center pixel beam profiles for both designs clearly show how diffracting apertures shield subsequent apertures. This is evident from the locations on the curves where sidelobe power suppression begins. A notable example is the suppression that begins at approximately $60^{\circ}$ which corresponds to the rejection angle $\theta_2$ of this design.

\begin{figure}[htbp]
\centering\includegraphics[width=0.7\textwidth]{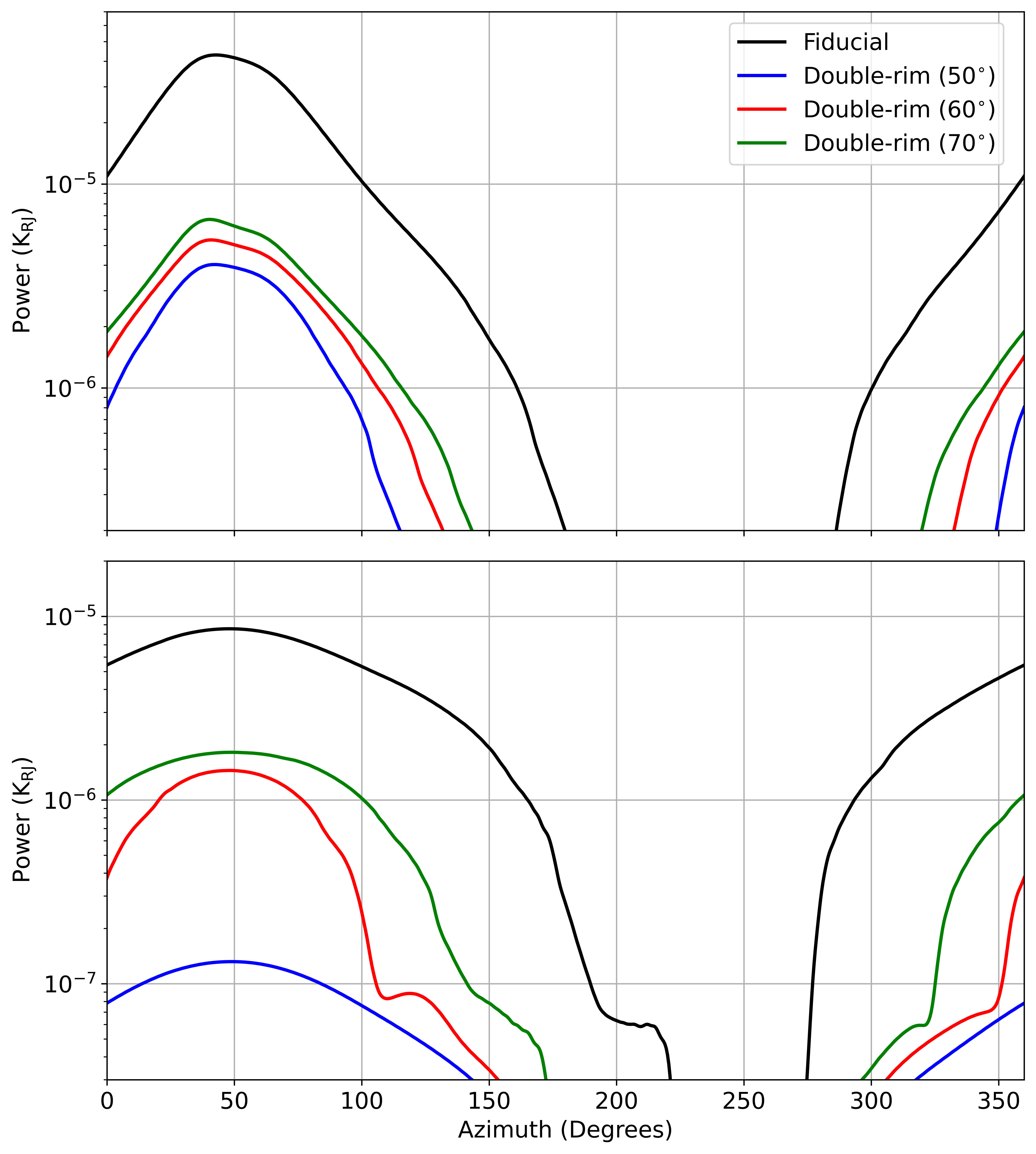}
\caption{Power pickup from beam sidelobes for different double-rim and fiducial forebaffle designs. Top: scan at 50$^{\circ}$ elevation; Bottom: scan at 65$^{\circ}$ elevation.}
\label{fig:sl_power}
\end{figure}

We calculate the total sidelobe power pickup in the Rayleigh-Jeans temperature units for a telescope observing near the slopes of Cerro Toco in the Atacama Desert. The simulation convolves the telescope's central detector beam with a detailed HEALPix\footnote{\url{https://healpix.sourceforge.io/}}~\cite{healpix} environment map. This map incorporates the sky modeled at 1 mm precipitable water vapor (PWV), a ground screen extending up to 5$^{\circ}$ elevation with a temperature equal to the sky at zenith, and the thermal emission from Cerro Toco modeled at 270 K temperature with fluctuations of 2$^{\circ}$ scale and 5 K RMS. The total sidelobe power pickup is determined by convolving this environment map with the beam map placed at a specific scanning coordinate
and subtracting a zero-level due to azimuthally symmetric power pick-up.
\par
Figure~\ref{fig:sl_power} demonstrates the gain in sidelobe power pickup by utilizing a double-rim forebaffle design in contrast to the fiducial design for scans at 50$^{\circ}$ and 65$^\circ$ elevation. The double-rim design consistently provides substantial suppression of the total power pickup across all azimuthal angles. Specifically, with the telescope observing along 50$^\circ$ elevation scans directly above the peak of Cerro Toco (azimuth of 50$^{\circ}$), the fiducial forebaffle design sees about an order of magnitude more sidelobe power than any of the double-rim designs. 
For 65$^\circ$ elevation scans, the double-rim forebaffle design with 50$^{\circ}$ rejection angle $\theta_2$ has almost two orders of magnitude less sidelobe pickup than the fiducial design with only a single traditional forebaffle.
This reduction directly translates into an improvement in observing efficiency. For instance, for a 50$^{\circ}$ elevation scan, if an experiment sets a maximum sidelobe power pickup requirement at 10$^{-6}$ K, the fiducial design restricts the allowable azimuthal scan range to between azimuths of $\sim$160$^{\circ}$ and $\sim$300$^{\circ}$, i.e., a total scan range of 140$^\circ$ in azimuth.  
In contrast, at the same observing elevation, the 50$^{\circ}$ and 60$^{\circ}$ rejection angle designs provide total scan ranges of $\sim$220$^{\circ}$ and $\sim$245$^{\circ}$ in azimuth, respectively. At 65$^{\circ}$ observing elevation, the corresponding azimuthal ranges allowed increase to $\sim$260$^{\circ}$ and $\sim$305$^{\circ}$, with no constraint  for the $\theta_2=50^{\circ}$ design.
We note that although these calculations have been performed for diffracted stray-radiation in the far sidelobes of the radiation pattern, significant improvements of all types of sources of straylight outlined in section~\ref{sec:problem} are expected with this design.

\section{Conclusion}
\label{sec:conclusion}

Stray radiation is a concern for sensitive astronomical observations of large angular scale emissions in the centimeter to sub-millimeter wavelength range. In this paper, we propose a double-rim forebaffle geometry that is expected to significantly reduce stray radiation contamination for such observations. Although we demonstrate the potential improvement for the specific case of observations from the Atacama plateau in Chile with a specific small refractive telescope geometry, the discussion is valid for observations with other instruments, in other sites, or even from space. The design is conceptually simple and relatively easy to implement in practice. For ground-based CMB observations, such a double-rim forebaffle, by allowing observations at lower elevation, with a wider azimuth range, or closer to the Sun, can potentially substantially improve the time efficiency and overall quality of the observations.

\acknowledgments
The authors thank LU Xuefeng and the IHEP AliCPT team for detailed dimensions of the AliCPT-1 forebaffle design, and colleagues at LBNL––Bobby Besuner, Shamik Ghosh, John Groh, Reijo Keskitalo, Adrian Lee, and Clara Vergès, for useful discussions. 
Some of the results in this paper have been derived using the healpy and HEALPix packages.

\bibliographystyle{JHEP}
\bibliography{biblio.bib}

\end{document}